\pdfoutput=1
\documentclass{optica-article}

\journal{opticajournal}

\articletype{Research Article}

\usepackage{amsmath,amsfonts}
\usepackage{graphicx}
\graphicspath{{figures/}}
\DefineJournalException{Journal of Lightwave Technology}{J. Lightwave Technol.}
\DefineJournalException{Scientific Reports}{Sci. Rep.}
\DefineJournalException{Royal Society Open Science}{R. Soc. Open Sci.}
\DefineJournalException{SIAM Journal on Optimization}{SIAM J. Optim.}
\DefineJournalException{Materials for Quantum Technology}{Mater. Quantum Technol.}
\DefineJournalException{Optics \& Laser Technology}{Opt. Laser Technol.}

\begin{document}

\title{Toward Topology-Optimized Foundry PDKs: A Seeded Design Framework for Multimode Interferometers}

\author{Jacob~M.~Hiesener,\authormark{1}
Archana~Kaushalram,\authormark{1}
Joshua~J.~Wong,\authormark{1}
Robert~P.~Pesch,\authormark{1}
and Stephen~E.~Ralph\authormark{1,*}}

\address{%
\authormark{1}School of Electrical and Computer Engineering, Georgia Institute of Technology, Atlanta, Georgia 30332, USA}

\email{\authormark{*}stephen.ralph@ece.gatech.edu}

\begin{abstract*}
We present an end-to-end design methodology for multimode interferometer (MMI)-based photonic devices that combines parameter optimization (PO) on analytical models with seeded topology optimization (TO) to maximize performance while preserving foundry design-rule compliance. A PO seed device is further refined via seeded TO, accessing a larger design space than analytical or parameterized methods alone can reach. We validate this pipeline on a 1$\times$2 splitter, a TE modal multiplexer, and a polarization splitter, with the first two fabricated on a commercial foundry process and measured. Seeded TO reduces the measured insertion loss of the 1$\times$2 splitter from 0.20 to 0.14 dB and improves the back-to-back TE\textsubscript{00} transmission of the modal multiplexer from $-$2.38 to $-$0.99 dB over 1274--1326 nm. Applying this pipeline to a commercial foundry process design kit (PDK)-provided 1$\times$2 and 2$\times$2 splitters improve measured transmission and bring the measured 2$\times$2 splitting ratio from 0.520 to 0.499, along with improved simulated fabrication robustness, offering a practical, foundry-validated route toward incorporating TO-designed components into commercial PDKs.
\end{abstract*}

\section{Introduction}

Commercial foundry platforms for silicon photonics enable scalable, high-volume fabrication of photonic integrated circuits (PICs) that integrate complex optical functionality using mature CMOS infrastructure \cite{liehr2020foundry,giewont2019300mm}. The passive routing and splitting components within foundry process design kits (PDKs) are typically designed using well-established analytical methods, such as multimode interferometer (MMI) theory and grating diffraction theory \cite{soldano1995optical, taillaert2006grating}. These methods yield devices with large feature sizes and simple geometries that are inherently robust to lithographic variation \cite{besse1994optical}, but whose performance is bounded by the assumptions of the underlying analytical model. Density-based topology optimization (TO) is a gradient-based inverse-design methodology that continuously evolves each voxel of a design region toward an optimal topology, unconstrained by such assumptions \cite{hammond2021photonic, hammond2022highperformance, piggott2017fabrication, molesky2018inverse, christiansen2021inverse, piggott2020inversedesigned}. TO has been demonstrated to generate high-performing solutions for design problems whose governing physics is complex or not well captured analytically, such as freeform metalenses and multi-port wavelength demultiplexers \cite{hammond2022highperformance,piggott2017fabrication,molesky2018inverse}; when the physics is well understood, as with MMI self-imaging \cite{soldano1995optical} or grating diffraction, analytical methods remain a strong starting point that TO can refine rather than replace \cite{hiesener2025seeded,hiesener2026enhancing}. Despite demonstrated performance advantages over conventionally designed devices, there are no reports of any TO-based photonic devices being assimilated into a commercial foundry PDK.

\par

Incorporating TO devices into a foundry PDK requires not just that they conform to stringent design rule checks (DRCs), which ensure accurate fabrication \cite{liehr2020foundry}, but that they remain robust to process variation to ensure high yield. DRC includes checks for minimum linewidth, linespacing, area, and enclosed area on each mask layer. In the TO framework, DRC compliance is enforced by augmenting the objective function with gradient-generating constraint functions that iteratively penalize design rule violations \cite{hammond2021photonic}. Because these constraint functions are linearly combined with the performance figure of merit (FOM), conflicting objectives (i.e., improving performance versus satisfying a linewidth or area constraint) can stall the optimization, preventing convergence to a high-performing, DRC-compliant design \cite{ballew2023constraining}.

\par

To overcome this limitation, we have developed seeded topology optimization (seeded TO): a modified TO implementation that enables effective optimization of commercial foundry-compatible integrated photonic devices \cite{hiesener2025seeded}. In seeded TO, a functional seed device—created by any design method, including parameter optimization (PO), shape optimization, or conventional TO—is iteratively blurred, corrected for DRC violations, and refined via TO. By initializing the optimization at a functional geometry that already conforms to design rules---rather than the random or gray-scale starting point typical of conventional TO---and by iteratively re-enforcing this compliance via a DRC fix step, seeded TO avoids the conflicting-objective stall and yields devices with improved fabrication robustness compared with traditional TO. Seeded TO has been demonstrated on both conventionally designed \cite{hiesener2026enhancing, hiesener2025seeded, hiesener2024topology} and inverse-designed \cite{hiesener2025seeded, hiesener2026seededgrating, hiesener2026tepass, agarwal2025inverse} seed devices.

\par

In this work, we present an end-to-end design methodology to create high-performing, foundry-compatible MMI-based photonic devices. MMI-based devices support a broad range of applications in integrated photonics, including power splitters, mode-control devices, and polarization-control devices. Power splitters divide optical power between multiple output ports, including N$\times$M splitters and arbitrary power splitters \cite{domenech2014mmi,hiesener2026enhancing,mizera2022polymer,michaels2020hierarchical}. Mode-control devices manipulate the modal content of light propagating in a waveguide, including modal multiplexers and mode converters \cite{chack2020broadband,gonzalezandrade2020experimental,uematsu2012design}. Polarization-control devices manipulate the polarization of light propagating in a waveguide, including polarization splitters, polarization rotators, and polarizers \cite{dai2013polarization,zafar2024recent,zhan2021pbs}. MMI-based structures are also used for WDM and filtering applications \cite{han2024wdm,hu2015angled}, though this work focuses on power splitters, mode-control devices, and polarization-control devices.

\par

In our design pipeline, we first design a parameterized model and perform a coarse PO step to determine the principal dimensions of an MMI seed structure \cite{bull2011convergence}. Seeded TO is then applied as a fine optimization step to maximize device performance while preserving fabrication robustness, in contrast to prior work that instead relies on shape optimization for this fine-optimization stage \cite{michaels2020hierarchical}. The three MMI-based devices we investigate share similar fundamental building blocks but realize distinct functionalities: a 1$\times$2 power splitter, a TE modal multiplexer, and a polarization splitter. The 1$\times$2 splitter and TE modal multiplexer are fabricated on a commercial foundry process and characterized to validate the proposed design pipeline. We then use the validated design pipeline to enhance the performance of two PDK-provided devices: a 1$\times$2 splitter and a 2$\times$2 splitter, with the goal of incorporating these devices in the GlobalFoundries (GF) Fotonix\textsuperscript{TM} PDK.

\section{Design Methods}

\begin{figure*}[!t]
\centering
\includegraphics[width=\textwidth]{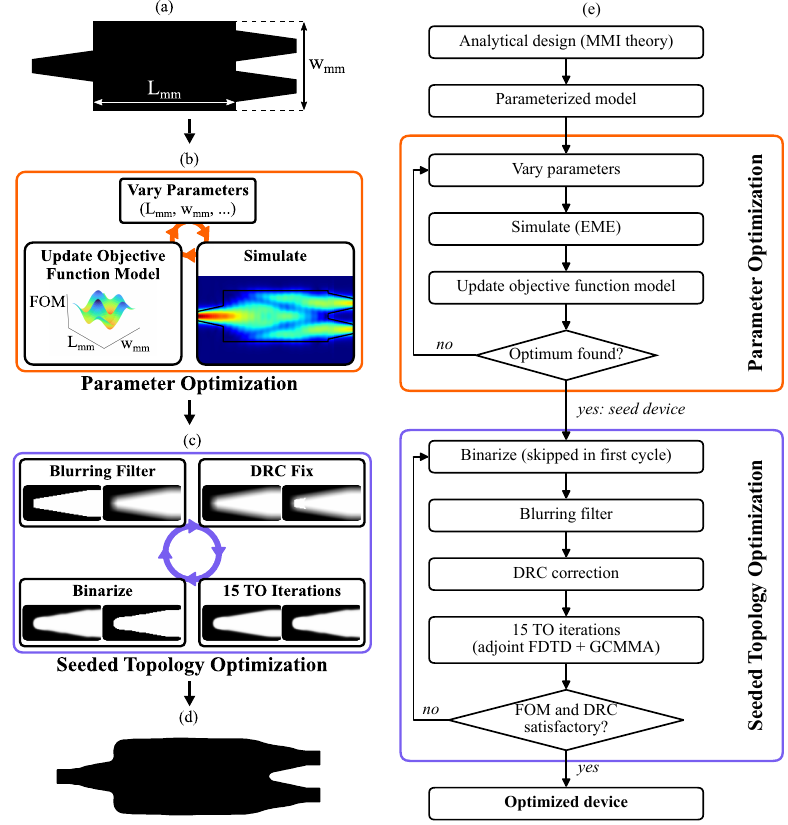}
\caption{Optimization pipeline for a 1$\times$2 MMI splitter. (a) A parameterized model of the MMI is created with defined features of the device, such as the multimode section length ($L_{mm}$) and width ($w_{mm}$). (b) The parameterized MMI is optimized using a surrogate objective function model where the MMI parameters are varied and the resulting device is simulated, computing the performance relative to a figure of merit (e.g., transmitted power). (c) The parameter-optimized device is then used as a seed for seeded TO. The seeded TO design loop consists of a binarization step, a blurring filter, a DRC correction algorithm, and 15 TO iterations, the combined effect of which is highlighted in the inset showing a zoomed-in view of the gap between the output tapers. (d) Multiple seeded TO cycles are performed to achieve an optimized device. (e) Flowchart of the design pipeline: the parameterized model is optimized via PO (orange) to generate the seed device, which is then refined over successive seeded TO cycles (purple) until the FOM and DRC compliance are satisfactory.}
\label{fig_sto_pipeline}
\end{figure*}

We now describe the multi-stage design methodology used throughout this work. A parameterized model of the device is first designed via analytical methods and then optimized using a surrogate objective function model; the resulting device is then used as a seed for seeded TO (Fig.~\ref{fig_sto_pipeline}).

\par

The self-imaging effect that governs MMI operation \cite{soldano1995optical} makes MMI devices an ideal starting point for a seeded TO-based design methodology. The search space for most MMIs is relatively small and well-behaved compared to the search space of traditional TO designs, which is often large and highly non-convex. By first finding a coarse optimum via PO on a parameterized MMI design, we can then exploit the local optimization capabilities of seeded TO to further improve device performance.

\subsection{Parameter Optimization}

MMI devices are traditionally designed using an analytical method and known MMI theory, where the dimensions of the multimode section are determined by the desired functionality of the device and the self-imaging effect \cite{soldano1995optical}. These parameters can then be optimized using heuristic methods, such as a grid search, or a surrogate objective function model \cite{bull2011convergence}. Because the mode is confined throughout the device, an eigenmode expansion (EME) method can be used to simulate the device and compute its performance relative to a figure of merit (FOM). We use the Lumerical EME solver \cite{lumerical} in this work; open-source alternatives such as MEOW, EMEpy, and CAMFR can also be used \cite{meow,hammond2022emepy,bienstman2001camfr}.

\par

Prior to PO, a parameterized model of the device is created in which the designer defines the parameters that describe the device geometry. For a simple 1$\times$2 splitter, for example, the designer may define the length and width of the multimode section as parameters to optimize ($L_{mm}$ and $w_{mm}$) while holding the taper widths, lengths, and locations constant (Fig.~\ref{fig_sto_pipeline}(a)). The designer then defines a FOM to optimize, such as the transmitted power through one or more ports of the device. The parameterized model is simulated with initial parameter values using an EME solver, and the FOM is computed; the parameter values are then varied and the device is re-simulated at another point in the design space, building the surrogate objective function model. This process repeats until the design space is sufficiently sampled and optimal parameter values are determined (Fig.~\ref{fig_sto_pipeline}(b)). The resulting PO device is then used as a seed for seeded TO.

\subsection{Seeded Topology Optimization}
\label{sec_seeded_to}

The details of seeded TO were first reported in \cite{hiesener2025seeded}; here, we provide a brief overview of the optimization process. The goal of seeded TO is to improve the performance of a device while maintaining or improving its compliance with foundry design rules. Seeded TO is a modified version of the density-based TO process presented in \cite{hammond2022highperformance}. Our density-based TO solver uses a hybrid time/frequency-domain adjoint-variable method configured to simultaneously optimize a user-specified FOM and the device's compliance with foundry design rules \cite{hammond2021photonic}. In each TO iteration, two or more finite-difference time-domain (FDTD) simulations are performed using the open-source Maxwell solver MEEP \cite{oskooi2010meep}. These consist of forward simulations, in which the device is excited with a source and the FOM is computed, and adjoint simulations, in which the device is excited by a time-reversed source at one of the output ports used to compute the FOM. These simulations are used to compute the gradient of the FOM with respect to the device permittivity distribution and backpropagated through the filtering and thresholding functions to determine the gradient with respect to the design parameters. This gradient is then used to update every design parameter via the globally convergent method of moving asymptotes (GCMMA) \cite{svanberg2002class}, implemented using the open-source NLopt package \cite{nlopt}, improving performance relative to the FOM.

\par

In seeded TO, the optimization is seeded with an already functional device geometry—a PO MMI-based device, in this work. The seeded TO process consists of a binarization step, a blurring filter, a DRC correction algorithm, and 15 TO iterations (Fig.~\ref{fig_sto_pipeline}(c)) \cite{hiesener2025seeded}. The binarization step maintains consistent blurring of the latent design parameters throughout seeded TO and is unnecessary for the first cycle, since the PO device is already binary. A user-defined blurring filter is then applied to the latent design parameters to enable perturbation of the device geometry during TO. The design parameters then undergo a DRC correction algorithm that expands or contracts violating regions of the device geometry, improving compliance with foundry design rules \cite{hiesener2025seeded}. TO iterations are then performed with constraint functions for the foundry design rules included, improving device performance while retaining or improving DRC compliance \cite{hammond2021photonic}. This process repeats over multiple seeded TO cycles until the device performance and DRC compliance are satisfactory (Fig.~\ref{fig_sto_pipeline}(d)). The complete design pipeline is summarized as a flowchart in Fig.~\ref{fig_sto_pipeline}(e). The resulting device can then be fabricated and tested to verify the performance of the design.

In addition to the hyperparameters of conventional TO, seeded TO introduces two sets of parameters: the blurring filter and the number of seeded TO cycles. The filter dimensions have a greater effect on device performance than the filter shape: a filter that is too large can obscure critical features of the seed, whereas a filter that is too small does not perturb the geometry sufficiently for TO to improve it. A rectangular averaging filter approximately one quarter of the single-mode waveguide width in each dimension is a practical starting point, refined by sweeping its dimensions \cite{hiesener2025seeded}. Other filter shapes (e.g., the donut filter in Section~\ref{sec_test_cases}) are occasionally tested for new design geometries \cite{wong2025seeding}. Each seeded TO cycle typically converges within 15 TO iterations, and the optimization generally plateaus after approximately four cycles, although additional cycles yield minor improvements for some devices. The filter dimensions and number of cycles used for each device are given in Sections~\ref{sec_test_cases} and \ref{sec_pdk_splitters}.

\section{Test Cases}
\label{sec_test_cases}

We demonstrate this design methodology on three test cases: a 1$\times$2 power splitter, a TE modal multiplexer, and a polarization splitter. Each test case begins as a parameterized model, is optimized via PO to generate a seed device, and is then further optimized using seeded TO. An objective function depending on the waveguide mode overlap is designed for each device, which is used for both the PO and seeded TO optimizations. The overlap coefficient is defined as,

\begin{equation}
a_m^{\pm} = c \int_A \left[ \mathbf{E}^*(r) \times \mathbf{H}_m^{\pm}(r) + \mathbf{E}_m^{\pm}(r) \times \mathbf{H}^*(r) \right] \cdot \hat{n} \, dA,
\label{eq_mode_overlap}
\end{equation}

where $a_m^{\pm}$ is the overlap coefficient of the $m$th mode for forward (+) and backward (--) directions, $\mathbf{E}(r)$ and $\mathbf{H}(r)$ are the Fourier-transformed total fields, $\mathbf{E}_m^{\pm}(r)$ and $\mathbf{H}_m^{\pm}(r)$ are the mode profiles for the forward and backward propagating modes, $A$ is the waveguide cross-sectional surface at the port of interest, and $c$ is the normalization constant, selected such that the overlap coefficient of the source mode is unity, i.e., $a_m^{\pm} = 1$. This is used to compute the transmitted power through a port of interest, from which the FOM for each test case is computed. 

\par 

Throughout this work, we optimize with a simulation resolution of 40 pixels/\textmu m and a design parameter resolution of 80 pixels/\textmu m, enabled by MEEP's subpixel smoothing functionality \cite{hammond2025unifying}. In the seeded TO optimization, the blurring filter used was a donut filter  defined by,

\begin{equation}
h(x,y) = G(x,y;\sigma_x,\sigma_y) - G\!\left(x,y;\tfrac{\sigma_x}{2},\tfrac{\sigma_y}{2}\right), \qquad G(x,y;\sigma_x,\sigma_y) = \exp\left[-\frac{1}{2}\left(\frac{x^2}{\sigma_x^2} + \frac{y^2}{\sigma_y^2}\right)\right],
\label{eq_donut_filter}
\end{equation}

where $G(x,y;\sigma_x,\sigma_y)$ is a 2D Gaussian with standard deviations $\sigma_x$ and $\sigma_y$ along the $x$- and $y$-directions, respectively, and $x$ and $y$ are design-region pixel coordinates centered on the filter kernel. The donut filter $h(x,y)$ is the difference between this Gaussian and a narrower Gaussian of half its standard deviation, producing a ring-shaped kernel that is normalized to unit sum before being applied. Rather than indiscriminately smoothing every length scale like a rectangular or Gaussian filter might, the donut filter selectively acts on features whose radius of curvature matches the ring radius, such as sharp corners and thin protrusions smaller than $\sigma$, concentrating the effect of seeded TO at transition regions like the taper-multimode interfaces.

\par

Each device was designed for the GF Fotonix\textsuperscript{TM} process in 160 nm thick silicon-on-insulator (SOI) for O-band wavelengths (1260--1360 nm). All simulations, in both the EME-based PO and the FDTD-based seeded TO, use the same three-dimensional layer stack constructed from publicly available values (a 160 nm silicon device layer with silicon dioxide cladding) and the built-in silicon and silicon dioxide material models of each solver. However, the proprietary foundry design rules are required to constrain the optimization to geometries that can be fabricated, allowing devices to be designed at the limits of the process. On-chip measured test structures were created using the GF Fotonix\textsuperscript{TM} PDK grating couplers. A standard fiber array setup was used to couple light into and out of the test structures, and transmission spectra were measured from 1270 to 1329 nm using a LUNA optical vector analyzer (OVA). The open-source software LabExT was used to automate the measurement process \cite{labext}. Grating coupler and waveguide insertion loss were calibrated out. Measured results are reported for four chips (1$\times$2 splitter) and five chips (TE modal multiplexer), and measured values are averaged in linear power across chips and over 1274--1326 nm.

\subsection{1$\times$2 Splitter}
\label{sec_1x2_splitter}

1$\times$2 splitters are fundamental low-loss routing components used throughout integrated photonic circuits. The splitter is a common test case for validating inverse design methods, with numerous works demonstrating ultra-compact TO splitters \cite{piggott2017fabrication,hammond2022highperformance,hansen2024inverse}. TO offers the greatest advantage over analytical design when aggressively minimizing footprint; at the larger, PDK-compatible scale considered in this work, analytical methods alone are already sufficient to generate a moderately performing seed device. The 1$\times$2 splitter designed in this work uses a conventional MMI geometry to divide optical power equally between two output waveguides; symmetry along the x-axis of the device is enforced in both the parameterized model and the seeded TO optimization to ensure an equal split. This simple 1$\times$2 splitter model is a common design approach used in many PDKs \cite{giewont2019300mm,liehr2020foundry} due to its compact footprint, high performance, and ease of fabrication \cite{besse1994optical,domenech2014mmi}. The FOM for the 1$\times$2 splitter is defined as,

\begin{equation}
f_1(\mathbf{E}) = 1 - 2|\alpha_0^{+}|^2,
\label{eq_1x2_splitter_fom}
\end{equation}

where $\alpha_0^{+}$ is the forward propagating fundamental mode coefficient at the output port of the top waveguide. A single overlap coefficient is sufficient to both minimize insertion loss and maintain an equal 50/50 split due to the enforced symmetry.

\par

\begin{figure}[!t]
\centering
\includegraphics[width=\textwidth]{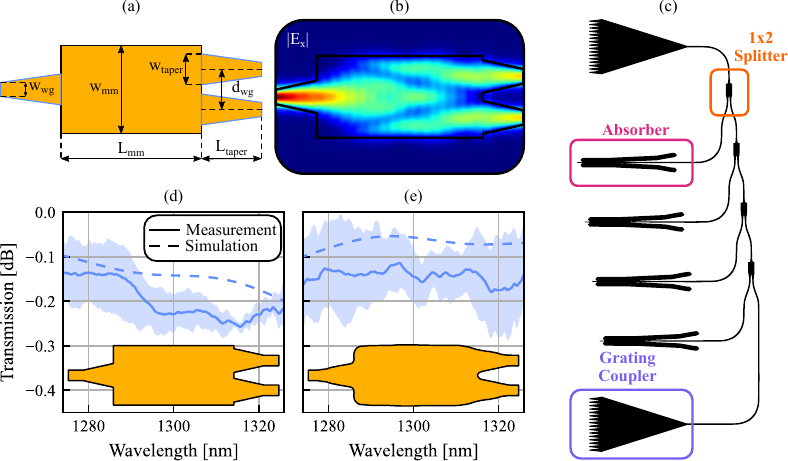}
\caption{(a) Parameterized geometry of a conventional 1$\times$2 MMI splitter used to generate the seed. Each of the blue outlined tapers has the same dimensions. (b) Electric field profile of the seed MMI splitter. (c) Test structure for measuring the performance of the MMI splitter via the cutback method (test structures included 1, 2, 3, or 4 cascaded MMIs). Simulated and measured transmission of the seed (d) and optimized (e) 1$\times$2 MMI splitter. The solid lines depict the mean (averaged in linear power) of the per-device transmission extracted from the cutback measurement of each chip; the light band depicts the minimum and maximum across the four measured chips at each wavelength.}
\label{fig_1x2_mmi}
\end{figure}

\begin{table}[h]
\centering
\caption{Parameterized geometry values for the 1$\times$2 MMI splitter seed device (Fig.~\ref{fig_1x2_mmi}(a)).}
\label{table_1x2_splitter_params}
{\renewcommand{\arraystretch}{1.0}
\begin{tabular}{lcc}
\hline
Parameter & Symbol & Value (\textmu m) \\
\hline
Waveguide width & $w_{wg}$ & 0.35 \\
Taper width & $w_{taper}$ & 0.8 \\
Taper length & $L_{taper}$ & 1.0 \\
Multimode section width & $w_{mm}$ & 2.0 \\
Multimode section length & $L_{mm}$ & 4.0 \\
Distance between output waveguide centers & $d_{wg}$ & 1.0 \\
\hline
\end{tabular}}
\end{table}

We set parameters for the waveguide width ($w_{wg}$), taper width ($w_{taper}$), taper length ($L_{taper}$), multimode section width ($w_{mm}$), multimode section length ($L_{mm}$), and distance between the center of the output waveguides ($d_{wg}$) (Fig.~\ref{fig_1x2_mmi}(a)). The standard single-mode waveguide width was used, and the taper dimensions were set to ensure efficient mode conversion into the multimode section. The multimode section width was set to 2.0 \textmu m to maintain a compact footprint while supporting multiple excited modes. The multimode section length and the distance between the output waveguides were then optimized via PO to minimize insertion loss; the resulting parameter values are given in Table~\ref{table_1x2_splitter_params}. The electric field profile of the seed device is shown in Fig.~\ref{fig_1x2_mmi}(b). The design space of the parameterized model is inherently limited, allowing PO to reach an optimum quickly; however, this optimum may fall short of what is achievable with the larger design space offered by a TO design method.

\par

Seeded TO was then applied using the previously described donut filter with $\sigma_x = \sigma_y = 5$ pixels (62.5 nm); four seeded TO cycles, comprising 60 total TO iterations, were performed to generate the optimized device, as no further performance gains were observed beyond this point. To measure the performance of both 1$\times$2 splitter variants, test structures were taped out and fabricated with 1, 2, 3, or 4 cascaded splitters at GlobalFoundries through a multi-project wafer (MPW) run on the GF Fotonix\textsuperscript{TM} platform and measured via the cutback method (Fig.~\ref{fig_1x2_mmi}(c)). The simulated and measured transmission spectra of the seed and optimized 1$\times$2 splitters are shown in Figs.~\ref{fig_1x2_mmi}(d) and \ref{fig_1x2_mmi}(e), respectively. The seeded TO variant shows significant improvement in both simulated and measured transmission relative to the seed device, achieving a mean measured (simulated) insertion loss of 0.14 (0.07) dB compared with 0.20 (0.14) dB for the seed over 1274--1326 nm. This performance is comparable to state-of-the-art 3~dB couplers designed via hierarchical shape optimization \cite{michaels2020hierarchical} in a more compact footprint validated with measured results.

\par

The performance of the PO device can be further improved by increasing the complexity of the parameterized model and the number of optimized parameters, at the cost of additional design complexity and longer optimization times. Increasing the multimode section width, and correspondingly its length, would excite additional modes, enabling finer engineering of the interference pattern via the taper geometry and potentially tighter mode confinement at the interference point.

\par

Similar to other seeded TO devices \cite{hiesener2025seeded,hiesener2026tepass,hiesener2024topology,hiesener2026seededgrating}, the sharp corners of the device are smoothed and minor perturbations are introduced throughout the structure by the seeded TO process. Part of the gap between the output tapers is filled in by the blurring filter; this is compensated for by a corresponding reduction in the multimode section width elsewhere in the structure. The most impactful adjustments occur at the tapers, which reshape the incident mode and, in turn, the modes excited in the multimode region. Having validated our design paradigm on this compact 1$\times$2 splitter, we next apply it to larger, more complex devices.

\subsection{TE Modal Multiplexer}

\begin{figure*}[!t]
\centering
\includegraphics[width=\textwidth]{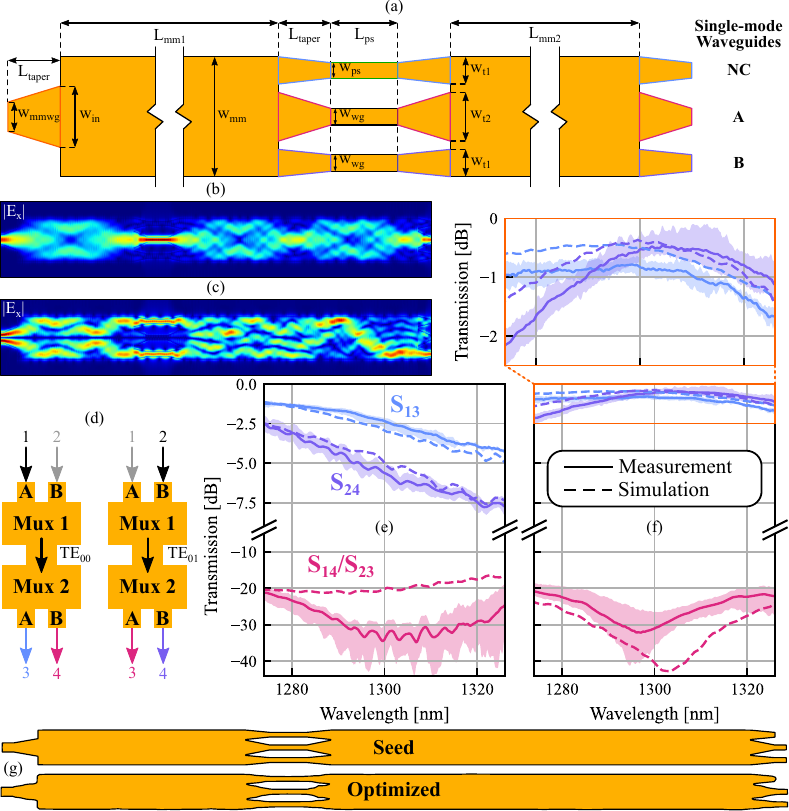}
\caption{(a) Parameterized geometry of the MMI-based TE modal multiplexer used to generate the seed, defining the taper, multimode-section, and single-mode waveguide ports. Tapers and waveguides outlined in the same color share identical dimensions. Simulated electric field profile of the seed mode multiplexer for TE\textsubscript{00} (b) and TE\textsubscript{01} (c) modes. (d) Test structure diagram for measuring modal multiplexer performance with input through port 1 (top) or port 2 (bottom). Simulated and measured S-parameters of the seed (e) and optimized (f) mode multiplexer. The solid lines depict the mean transmission (averaged in linear power) across the five measured chips; the light band depicts the minimum and maximum across those chips at each wavelength (S\textsubscript{14} and S\textsubscript{23} are pooled for the crosstalk). (g) To-scale geometry of the seed and optimized modal multiplexer.}
\label{fig_mode_mux}
\end{figure*}

Modal multiplexers and mode converters are essential building blocks for any multimode photonic system, enabling applications such as mode-division multiplexing and modal switching. Here we design a TE modal multiplexer that converts the fundamental mode (TE\textsubscript{00}) of two separate single-mode waveguides into the first two TE modes of a multimode waveguide (TE\textsubscript{00} and TE\textsubscript{01}). This device (Fig.~\ref{fig_mode_mux}(a)) is based on the two-MMI, phase-shifter modal multiplexer design of Uematsu \textit{et al.} \cite{uematsu2012design} and, as a result, is much larger than previous modal multiplexers designed using TO \cite{frellsen2016topology,hiesener2025seeded}. The device cascades a 1$\times$3 MMI splitter, a phase shifter, and a 3$\times$3 MMI splitter, with the third port of each MMI left unconnected: the first MMI splits each single-mode input across the two connected waveguides, the phase shifter imparts a relative phase shift between them, and the second MMI recombines the phase-shifted fields such that constructive and destructive interference route each input condition to the corresponding mode of the output multimode waveguide. This device requires parallel optimization of both the TE\textsubscript{00} ($f_1(\mathbf{E})$) and TE\textsubscript{01} ($f_2(\mathbf{E})$) input conditions, with FOMs defined as \cite{hiesener2025seeded},

\begin{equation}
f_1(\mathbf{E}) = 1 - |\alpha_{0,A}^{+}|^2 + b|\alpha_{0,B}^{+}|^2, \quad f_2(\mathbf{E}) = 1 - |\alpha_{0,B}^{+}|^2 + b|\alpha_{0,A}^{+}|^2,
\label{eq_mode_mux_fom}
\end{equation}

where $b$ is the user-specified extinction coefficient, $\alpha_{0,A}^{+}$ is the forward propagating fundamental mode coefficient of single-mode waveguide A (port 1/4), and $\alpha_{0,B}^{+}$ is the forward propagating fundamental mode coefficient of single-mode waveguide B (port 2/3). This formulation is designed to both maximize transmission and maximize the extinction ratio (ER).

\par 

For this TE modal multiplexer we define parameters for the input and output waveguides ($w_{wg}$, $w_{mmwg}$), taper shapes ($w_{in}$, $L_{taper}$, $w_{t1}$, $w_{t2}$), multimode section dimensions ($w_{mm}$, $L_{mm1}$, $L_{mm2}$), and the phase-shift section ($w_{ps}$, $L_{ps}$), as shown in Fig.~\ref{fig_mode_mux}(a) and Table~\ref{table_mode_mux_params}. To ensure compatibility with other devices, the standard single-mode waveguide width ($w_{wg}$) is used; the multimode waveguide width ($w_{mmwg}$) is set to 0.7 \textmu m to support only the first two TE modes.

\par

Keeping the device length within a regime that could be simulated in a manageable time without excessive numerical dispersion error using an FDTD solver \cite{hammond2022highperformance,teixeira2023finitedifference,abusamak2016comprehensive} was an additional design consideration, since this is the longest device (64 \textmu m) optimized using our TO framework to date. This constraint guided our choice to fix the taper length ($L_{taper}$ = 2.0 \textmu m), length of the phase-shift waveguides ($L_{ps}$ = 3.0 \textmu m), multimode section width ($w_{mm}$ = 2.8 \textmu m), and input taper width ($w_{in}$ = 1.4 \textmu m) (Table~\ref{table_mode_mux_params}). The output taper widths were correlated so that only one was optimized,

\begin{equation}
    w_{t1} = \frac{w_{mm} - w_{t2} - 2 \times 0.1}{2},
\label{eq_mode_mux_taper_correl}
\end{equation}

where 0.1 \textmu m is the gap between the output tapers. Phase-shift waveguide width ($w_{ps}$), inner output taper width ($w_{t2}$), and the two multimode section lengths ($L_{mm1}$, $L_{mm2}$) were optimized via PO. The simulated fields from the EME solver for the PO variant are shown in Figs.~\ref{fig_mode_mux}(b) and \ref{fig_mode_mux}(c). 

\par

\begin{table}[h]
\centering
\caption{Parameterized geometry values for the TE modal multiplexer seed device (Fig.~\ref{fig_mode_mux}(a)).}
\label{table_mode_mux_params}
{\renewcommand{\arraystretch}{1.0}
\begin{tabular}{lcc}
\hline
Parameter & Symbol & Value (\textmu m) \\
\hline
Single-mode waveguide width & $w_{wg}$ & 0.35 \\
Multimode waveguide width & $w_{mmwg}$ & 0.7 \\
Input taper width & $w_{in}$ & 1.4 \\
Taper length & $L_{taper}$ & 2.0 \\
Phase-shift waveguide width & $w_{ps}$ & 0.33 \\
Phase-shift waveguide length & $L_{ps}$ & 3.0 \\
Outer output taper width & $w_{t1}$ & 0.725 \\
Inner output taper width & $w_{t2}$ & 1.150 \\
Multimode section width & $w_{mm}$ & 2.8 \\
First multimode section length & $L_{mm1}$ & 16.92 \\
Second multimode section length & $L_{mm2}$ & 34.24 \\
\hline
\end{tabular}}
\end{table}

Seeded TO was then applied using the donut blurring filter with $\sigma_x = \sigma_y = 4$ pixels (50 nm). The optimization was parallelized across 10 dual-socket nodes, each with two Intel Xeon Gold 6226 CPUs, on the Georgia Tech PACE compute cluster \cite{pace}. Owing to the length of the device (64 \textmu m), each TO iteration took between 1 and 2 hours; three seeded TO cycles, comprising 45 total TO iterations, were performed to generate the seeded TO variant. To measure device performance, a back-to-back test structure was taped out and fabricated at GlobalFoundries through an MPW run on the GF Fotonix\textsuperscript{TM} platform, and its S-parameters were measured, following the same approach used for other TE modal (de)multiplexers (Fig.~\ref{fig_mode_mux}(d)) \cite{hiesener2025seeded,gonzalezandrade2020experimental}.

\par

The simulated and measured S-parameters for the PO and seeded TO TE modal multiplexer variants are shown in Figs.~\ref{fig_mode_mux}(e) and \ref{fig_mode_mux}(f), respectively. The S\textsubscript{13} transmission encapsulates two passes through the device for the TE\textsubscript{00} channel, while the S\textsubscript{24} transmission encapsulates two passes for the TE\textsubscript{01} channel. The S\textsubscript{14} and S\textsubscript{23} transmissions capture the crosstalk between the two channels; because the back-to-back test structure consists of two identical, mirrored modal multiplexers, reciprocity ensures these crosstalk terms are equal. 

\par

Seeded TO significantly improves the conversion efficiency of both modes (Figs.~\ref{fig_mode_mux}(e) and \ref{fig_mode_mux}(f)). Over 1274--1326 nm, the seeded TO variant achieves a mean measured (simulated) transmission of $-$0.99 ($-$0.67) dB for the TE\textsubscript{00} channel and $-$0.91 ($-$0.75) dB for the TE\textsubscript{01} channel, with crosstalk of $-$24.5 ($-$28.9) dB, compared with $-$2.38 ($-$2.75) dB, $-$5.11 ($-$4.69) dB, and $-$27.3 ($-$19.4) dB, respectively, for the seed. The performance of the seeded TO variant significantly exceeds previously designed TO \cite{frellsen2016topology,hiesener2025seeded} and parameter-optimized \cite{uematsu2012design,chack2020broadband,gonzalezandrade2020experimental,guo2017silicon} TE modal multiplexers.

\par

Compared with simulation, the measured crosstalk of the seed is approximately 8 dB lower, whereas that of the seeded TO variant is approximately 4.5 dB higher. The S\textsubscript{14} and S\textsubscript{23} crosstalk of the back-to-back test structure is the coherent sum of the contributions from both modal multiplexers; depending on their relative phase, the measured crosstalk can be higher or lower than that of a single device. In addition, fabrication variations such as silicon thickness and waveguide width variations, non-vertical sidewalls, and lithographic corner rounding that is only partially compensated by optical proximity correction cause the fabricated device to deviate from the simulated geometry; the phase-shift section of this device is particularly sensitive to such deviations \cite{besse1994optical,lu2017performance}. The measured crosstalk traces lie well above the measurement noise floor, and S\textsubscript{14} and S\textsubscript{23} agree to within 0.7 dB, indicating that these differences reflect device behavior rather than measurement limitations. Despite these discrepancies, the mean measured crosstalk of both variants remains below $-$24 dB, while seeded TO substantially improves the transmission of both channels.

\par

As with the 1$\times$2 splitter, the most impactful adjustments occur at the tapers, which are reshaped to excite slightly different modes in the multimode section (Fig.~\ref{fig_mode_mux}(g)). Scaling this design methodology to structures larger than this device will likely require alternative simulation and optimization frameworks such that optimization is feasible. One such approach is to divide the device into sections, such as the two MMI stages and the phase-shift section of this modal multiplexer, and optimize each section in parallel within its own, smaller FDTD simulation. The resulting three simulation domains would speed up the optimization, while the full-device performance could still be optimized at each iteration by cascading the section scattering matrices. However, sectional optimization may neglect reflections and higher-order modes coupled between sections, potentially reducing performance relative to global optimization. Seeded TO is well suited to such a hybrid scheme, in which sectionally optimized devices serve as the seed for a final global seeded TO stage that recovers the interface effects. 

\subsection{Polarization Splitter}
\label{sec_pol_splitter}

\begin{figure*}[!t]
\centering
\includegraphics[width=\textwidth]{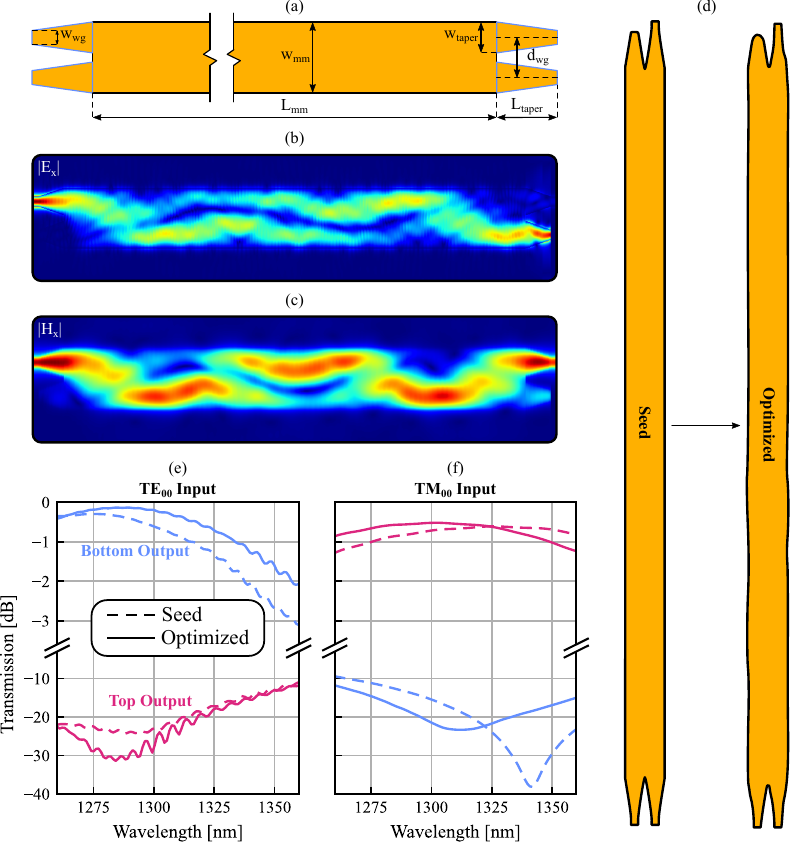}
\caption{(a) Parameterized geometry of the seed MMI-based polarization splitter, defining the dimensions of the multimode section and output tapers. Simulated electric field profiles of the seed polarization splitter for TE\textsubscript{00} (b) and TM\textsubscript{00} (c) inputs. (d) To-scale geometry of the seed and optimized polarization splitter. Simulated transmission of the seed and optimized polarization splitter for TE\textsubscript{00} (e) and TM\textsubscript{00} (f) inputs; dashed lines depict the seed device, solid lines depict the device optimized via seeded TO.}
\label{fig_pol_splitter}
\end{figure*}

On-chip polarization control provides an additional degree of freedom for multiplexing and routing signals in integrated photonic systems \cite{dai2013polarization,zafar2024recent}. A polarization splitter separates the fundamental TE and TM modes of a single-mode waveguide into two separate single-mode waveguides. The seed device uses a 2$\times$2 MMI configuration with a multimode section length and width chosen such that the TE and TM self-imaging distances coincide, with the TE mode imaging to the bottom waveguide and the TM mode to the top waveguide (Figs.~\ref{fig_pol_splitter}(a)--\ref{fig_pol_splitter}(c)). Like the TE modal multiplexer, this device requires parallel optimization of both the TE\textsubscript{00} ($f_1(\mathbf{E})$) and TM\textsubscript{00} ($f_2(\mathbf{E})$) input conditions, with FOMs defined as,

\begin{equation}
f_1(\mathbf{E}) = 1 - |\alpha_{0,bot}^{+}|^2, \quad f_2(\mathbf{E}) = 1 - |\alpha_{1,top}^{+}|^2,
\label{eq_pol_splitter_fom}
\end{equation}

where $\alpha_{0,bot}^{+}$ is the forward propagating TE\textsubscript{00} mode coefficient of the bottom output waveguide, and $\alpha_{1,top}^{+}$ is the forward propagating TM\textsubscript{00} mode coefficient of the top output waveguide. These FOMs maximize the transmission of each mode into its respective output waveguide; extinction of the undesired mode at each port is implicitly enforced.

\par

For the polarization splitter parameterized model, we define parameters for the waveguide size and position ($w_{wg}$, $d_{wg}$), taper dimensions ($w_{taper}$, $L_{taper}$), and multimode section dimensions ($w_{mm}$, $L_{mm}$), as shown in Fig.~\ref{fig_pol_splitter}(a) and Table~\ref{table_pol_splitter_params}. The waveguide width is set to the standard single-mode waveguide width for O-band wavelengths on this platform, and the taper length is set to 2 \textmu m to ensure efficient conversion between the waveguide mode and the multimode section without making the device excessively long. The taper width and distance between the output waveguide centers are both set as functions of the multimode section width,

\begin{equation}
w_{taper} = \frac{w_{mm} - 0.1}{2}, \quad d_{wg} = \frac{w_{mm} + 0.1}{2},
\label{eq_pol_spltr_correl}
\end{equation}

where 0.1 \textmu m is the gap between the output tapers. These parameter definitions reduce the design space to just the multimode section length and width, which are optimized via PO (Table~\ref{table_pol_splitter_params}). Seeded TO was then applied using the previously described donut filter with $\sigma_x = \sigma_y = 4$ pixels (50 nm); 10 seeded TO cycles, comprising 150 total TO iterations, were performed to generate the optimized structure shown in Fig.~\ref{fig_pol_splitter}(d).

\par

\begin{table}[h]
\centering
\caption{Parameterized geometry values for the polarization splitter seed device (Fig.~\ref{fig_pol_splitter}(a)).}
\label{table_pol_splitter_params}
{\renewcommand{\arraystretch}{1.0}
\begin{tabular}{lcc}
\hline
Parameter & Symbol & Value (\textmu m) \\
\hline
Waveguide width & $w_{wg}$ & 0.35 \\
Taper width & $w_{taper}$ & 1.0 \\
Taper length & $L_{taper}$ & 2.0 \\
Multimode section width & $w_{mm}$ & 2.1 \\
Multimode section length & $L_{mm}$ & 37.9 \\
Distance between output waveguide centers & $d_{wg}$ & 1.1 \\
\hline
\end{tabular}}
\end{table}

For the TE\textsubscript{00} input, the PO seed exhibits low crosstalk, but its transmission rolls off toward the long-wavelength edge of the O-band (Fig.~\ref{fig_pol_splitter}(e)). Averaged over the O-band (1260--1360 nm), seeded TO improves the TE\textsubscript{00} transmission from $-$1.08 to $-$0.63 dB (worst case $-$3.11 to $-$2.09 dB) by shifting the response toward the center of the optimization band, while the TE\textsubscript{00} crosstalk remains roughly the same ($-$17.1 and $-$18.1 dB; worst case $-$11.0 and $-$11.8 dB). The TM\textsubscript{00} response of the PO seed is already near-optimal (Fig.~\ref{fig_pol_splitter}(f)); seeded TO slightly improves the O-band-averaged TM\textsubscript{00} transmission from $-$0.78 to $-$0.71 dB (worst case $-$1.28 to $-$1.24 dB) and reduces the O-band-averaged TM\textsubscript{00} crosstalk from $-$15.1 to $-$17.2 dB (worst case $-$9.4 to $-$11.8 dB).

\par

Similar to the previous MMI-based devices, seeded TO primarily perturbs the taper regions of the polarization splitter. These regions set the amplitude and phase of the excited modes in the multimode region, as well as the scattering and reflection at the transition between the taper and the multimode section. By reshaping the tapers and interfaces, seeded TO adjusts the modal excitation of the multimode section and reduces interface losses, sharpening the self-image at the output tapers with only minor perturbations to the multimode section, which preserve the principal dimensions of the seed device. For a relatively simple device such as an MMI, shape optimization of a parameterized taper and multimode geometry can achieve comparable performance \cite{michaels2020hierarchical}; the greater design freedom of seeded TO becomes increasingly valuable for components with stricter performance requirements or more complex geometries. Unlike shape optimization, which is restricted to the design space spanned by a predefined parameterization, seeded TO allows every pixel of the design region to evolve, including changes in topology, without requiring the designer to construct a parameterization of the features to be optimized. In addition, seeded TO tends to remove extraneous features, yielding simpler geometries whose measured performance closely matches simulation \cite{wong2025seeding,hiesener2025seeded}. Furthermore, seeded TO makes additional modifications to the multimode section of the polarization splitter at a few locations where the interference patterns of the two polarizations diverge, affecting the performance of one polarization without disturbing the other. These examples demonstrate the capability of the seeded TO design paradigm across routing, mode-control, and polarization-control MMI-based devices.

\section{Toward Seeded TO PDK Splitters}
\label{sec_pdk_splitters}

Many PDK devices are MMI-based designs, as the self-imaging effect that governs MMI operation is inherently broadband and tolerant to the lithographic variation encountered in foundry fabrication \cite{soldano1995optical,besse1994optical}. The previous section demonstrates the capability of seeded TO to improve the performance of custom-designed MMI-based devices fabricated on a commercial foundry platform. Here we apply seeded TO to two PDK-provided devices, a 1$\times$2 splitter and a 2$\times$2 splitter \cite{giewont2019300mm}. Both the PDK and seeded TO variants of each splitter were taped out and fabricated at GlobalFoundries through an MPW run on the GF Fotonix\textsuperscript{TM} platform and measured on five chips from two wafers using the setup described in Section~\ref{sec_test_cases}. The per-device transmission of each splitter was extracted from cutback structures with 4, 8, and 16 cascaded devices; for the 2$\times$2 splitter, this is the transmission of the bright output path in the cascade, which also includes the penalty of any splitting imbalance. The splitting ratio was measured on single devices, with each output normalized to an on-chip grating coupler loopback.

\par

Instead of the donut filter used for blurring the test cases, a rectangular averaging filter was used for these devices,

\begin{equation}
h(x,y) = \frac{1}{n_x n_y}, \qquad -\frac{n_x-1}{2} \leq x \leq \frac{n_x-1}{2}, \quad -\frac{n_y-1}{2} \leq y \leq \frac{n_y-1}{2},
\label{eq_rect_filter}
\end{equation}

where $n_x$ and $n_y$ are the filter width and height, in design-region pixels, along the $x$- and $y$-directions, respectively, and $x$ and $y$ are design-region pixel coordinates centered on the filter kernel, as in Eq.~(\ref{eq_donut_filter}). Unlike the donut filter, this rectangular averaging filter applies a uniform weight over its entire footprint, producing a low-pass blur rather than concentrating the perturbation at a specific radius of curvature. A small $n_x$ is used to minimize blurring in the $x$-direction, since a large $x$-blur can unintentionally fill the gap between adjacent input or output tapers, extending the multimode section and degrading performance. This asymmetric blur mitigates that effect, enabling more seeded TO cycles before the taper gap significantly closes. The PDK splitters were optimized with an alternate version of our seeded TO implementation that supports this filter. In a previous study comparing rectangular, Gaussian, and donut filters of various sizes, we found that the filter dimensions affect device performance more than the filter type, and that seed geometries without disjoint features perform well with small rectangular filters \cite{wong2025seeding}, including the MMI-based seeds in this work. The filter is applied only once per seeded TO cycle and its cost is negligible compared with the FDTD simulations; therefore, the choice of filter does not change the simulation time per iteration.

\begin{figure*}[!t]
\centering
\includegraphics[width=\textwidth]{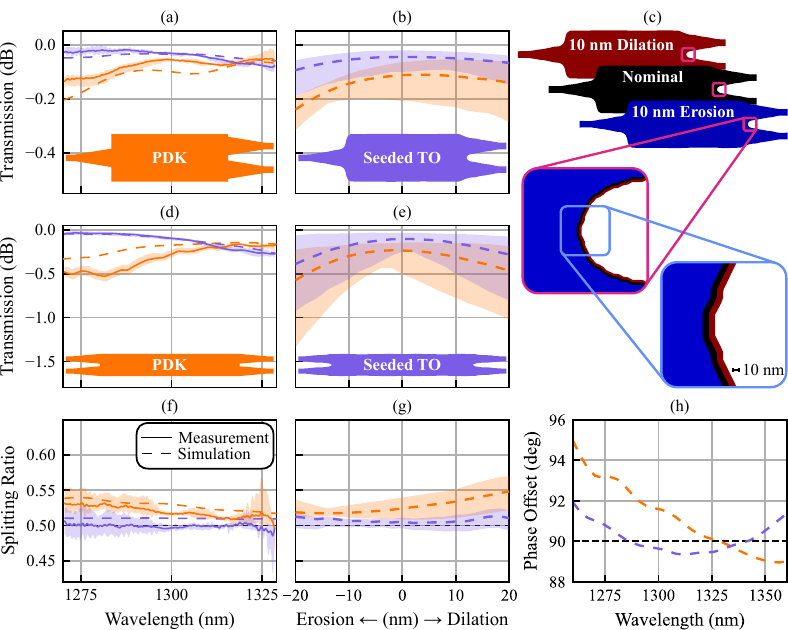}
\caption{(a) Measured (solid) and simulated (dashed) transmission spectra (sum of power in both output waveguides) for the PDK and seeded TO 1$\times$2 splitter variants. (b) Over-/under-etch study showing the transmission of eroded (over-etched) and dilated (under-etched) versions of each 1$\times$2 splitter over O-band. (c) To-scale overlay of the dilated (red), original (black), and eroded (blue) seeded TO 1$\times$2 splitter geometries for an illustrative $\pm$10 nm dilation/erosion, with successive zoom-ins on the gap between the output tapers. (d) Measured (solid) and simulated (dashed) transmission spectra for the PDK and seeded TO 2$\times$2 splitters. (e) Over-/under-etch study showing the transmission of eroded and dilated versions of each 2$\times$2 splitter over O-band. (f) Measured (solid) and simulated (dashed) splitting ratio spectra for each 2$\times$2 splitter. (g) Over-/under-etch study showing the splitting ratio for each splitter. (h) Simulated phase offset between the two output waveguides of each 2$\times$2 splitter variant relative to the ideal 90\textdegree{} quadrature condition (dashed line), over O-band. The solid lines in (a), (d), and (f) depict the mean in linear power across the measured chips; the light band depicts the minimum and maximum across those chips at each wavelength. The dashed lines in (b), (e), and (g) depict the simulated O-band average at each dilation; the light band depicts the minimum and maximum over O-band (the black dashed line in (g) marks the ideal 0.5 splitting ratio).}
\label{fig_pdk_splitters}
\end{figure*} 

\subsection{1$\times$2 MMI Splitter}

The 1$\times$2 splitter provided in the GF Fotonix\textsuperscript{TM} PDK is slightly larger than the custom-designed 1$\times$2 splitter demonstrated in Section~\ref{sec_1x2_splitter}. This variant uses the same conventional MMI geometry as the custom-designed device to divide the input power equally between two output waveguides. Because the underlying geometry and functionality match, seeded TO is applied using the same FOM (Eq.~(\ref{eq_1x2_splitter_fom})) described in Section~\ref{sec_1x2_splitter}. The filter width was fixed at $n_x = 2$ pixels (25 nm) to minimize blurring along the $x$-direction, and the filter height was swept from $n_y = 9$ to 21 pixels in steps of 2 pixels in separate optimizations. The best-performing optimization after the first few seeded TO cycles was selected, corresponding to $n_y = 13$ pixels (162.5 nm). Ten seeded TO cycles, comprising 150 total TO iterations, were performed to generate the optimized device.

\par

Similar to the other MMI-based devices, seeded TO primarily perturbs the taper regions and the taper-multimode section interface (Fig.~\ref{fig_pdk_splitters}(a)). Seeded TO introduces slight ripples along the tapers and smooths the sharp corners of the PDK variant's multimode section. These modifications improve the measured (simulated) transmission, averaged over 1274--1326 nm, from $-$0.08 ($-$0.10) dB for the PDK variant to $-$0.04 ($-$0.04) dB for the seeded TO variant (Fig.~\ref{fig_pdk_splitters}(a)) and shift the operating wavelength closer to the center of O-band. The measured and simulated transmission agree to within 0.03 dB for both variants, indicating that the simulations accurately predict the performance of the fabricated devices.

\par

To further probe the fabrication robustness of the seeded TO variant, we performed an over-/under-etch study, eroding and dilating each MMI geometry and re-simulating its performance over O-band to emulate lithographic variation (Figs.~\ref{fig_pdk_splitters}(b) and \ref{fig_pdk_splitters}(c)). Simulations were performed directly on the exported, interpolated GDS geometry at a resolution of 80 pixels/\textmu m using the subpixel-smoothed projection technique \cite{hammond2025unifying}, which enables accurate simulation of curved and angled interfaces on a Cartesian grid. The seeded TO variant significantly outperforms the PDK splitter under under-etch (dilation) and slightly outperforms it under over-etch (erosion). This robustness is consistent with the measured improvement of the fabricated seeded TO variant.

\subsection{2$\times$2 MMI Splitter}

A 2$\times$2 splitter poses a more complex design problem than a 1$\times$2 splitter, as its offset input waveguides preclude an implicit 0.5 splitting ratio. The GF Fotonix\textsuperscript{TM} PDK 2$\times$2 splitter shares its geometry with the polarization splitter described in Section~\ref{sec_pol_splitter}, but with smaller dimensions and a shorter multimode section so that the output tapers begin where the fields in the multimode section are evenly split. Like the polarization splitter, this device requires parallel optimization of both output waveguides, with FOMs defined as,

\begin{equation}
f_1(\mathbf{E}) = | 1 - 2|\alpha_{0,bot}^{+}|^2 |, \quad f_2(\mathbf{E}) = | 1 - 2|\alpha_{0,top}^{+}|^2 |,
\label{eq_2x2_splitter_fom}
\end{equation}

where $\alpha_{0,bot}^{+}$ is the forward propagating fundamental mode coefficient of the bottom output waveguide, and $\alpha_{0,top}^{+}$ is the forward propagating fundamental mode coefficient of the top output waveguide. Separating the FOM by output waveguide, rather than combining them into a single objective, allows the minimax formulation of our TO implementation \cite{hammond2021photonic} to simultaneously maximize transmission and drive the splitting ratio toward 0.5. For this device, we define the splitting ratio as,

\begin{equation}
\text{Splitting Ratio} = \frac{|\alpha_{0,bot}^{+}|^2}{|\alpha_{0,bot}^{+}|^2 + |\alpha_{0,top}^{+}|^2},
\label{eq_2x2_splitting_ratio}
\end{equation}

the fraction of the total transmitted power routed to the bottom output waveguide. 

\par

Symmetry along both the x- and y-axes of the device was enforced throughout the optimization, preserving structural symmetry between the two output ports. The same filter sweep was performed as for the 1$\times$2 PDK splitter ($n_x = 2$ pixels; $n_y$ from 9 to 21 pixels in steps of 2 pixels), and the best-performing optimization after the first few seeded TO cycles corresponded to $n_y = 15$ pixels (187.5 nm). Four seeded TO cycles, comprising 60 total TO iterations, were performed to generate the optimized device.

\par

In addition to perturbing the tapers and taper-multimode section interface, seeded TO reshapes the multimode section, redistributing the field before the output tapers to achieve a more even power split between the output ports. These modifications improve the measured (simulated) transmission, averaged over 1274--1326 nm, from $-$0.32 ($-$0.19) dB to $-$0.12 ($-$0.12) dB (Fig.~\ref{fig_pdk_splitters}(d)), and simulated over-/under-etch studies show that the transmission of the seeded TO variant remains superior across all erosions and dilations (Fig.~\ref{fig_pdk_splitters}(e)); more importantly, they significantly improve the measured (simulated) splitting ratio, averaged over the same band, from 0.520 (0.528) to 0.499 (0.510), with the mean measured splitting ratio of the seeded TO variant remaining within 0.008 of 0.5 across the measurement band (Fig.~\ref{fig_pdk_splitters}(f)). The seeded TO splitting ratio is also significantly more robust to fabrication variation than that of the PDK device (Fig.~\ref{fig_pdk_splitters}(g)). The measured transmission of the seeded TO variant decreases toward longer wavelengths and falls below that of the PDK variant above approximately 1313 nm, a roll-off that is reproduced by simulation of the fabricated layout.

\par

The measured transmission and splitting ratio for the seeded TO variant agree with simulation to within 0.01 dB and 0.011, respectively, and the measured splitting ratio of the PDK variant agrees with simulation to within 0.008. Differences this small approach the precision of the comparison, which is limited by the accuracy of the simulations and of the cutback extraction, as well as by fabrication variations across the wafer. The larger transmission discrepancy for the PDK variant arises partly because the cutback measurement of the transmission includes the penalty of its splitting imbalance, which is not captured by the simulated single-device transmission. In addition, the seeded TO variant maintains a simulated phase offset between the two output waveguides that is closer to the ideal 90\textdegree{} quadrature condition (Fig.~\ref{fig_pdk_splitters}(h)); this is important for applications such as Mach--Zehnder interferometers and coherent receivers.  The measured improvements in transmission and splitting ratio support the seeded TO variants as potential replacements for the PDK variants.

\section{Conclusions}

Intelligent application of inverse design methods is critical to achieving high-performing photonic devices. When a robust analytical model already exists, full, unconstrained TO is often unnecessary: the model already provides most of the achievable performance, and the remaining gains are better captured by refining that seed than by re-deriving the design from scratch. Conventional TO instead remains best suited to problems that lack such a starting point, or that demand a degree of compactness beyond what any analytical geometry can offer. For the common case of an established analytical model, such as MMI theory, an end-to-end pipeline that combines analytical and inverse design methods, like the one presented here, should be used instead. Our approach consists of designing a parameterized model, optimizing that model via PO, and using seeded TO to refine the design. A target use case of this design paradigm is enhancing commercial foundry PDK components. To validate this design methodology, we designed a 1$\times$2 splitter, a TE modal multiplexer, and a polarization splitter. The 1$\times$2 splitter and TE modal multiplexer were fabricated on the GF Fotonix\textsuperscript{TM} process and subsequently measured.

\par

The measured (simulated) insertion loss of the 1$\times$2 splitter improved from 0.20 (0.14) dB for the seed to 0.14 (0.07) dB via seeded TO, comparable to state-of-the-art 3~dB couplers designed via hierarchical shape optimization \cite{michaels2020hierarchical} in a more compact footprint on a 160 nm SOI platform. The per-channel transmission of the TE modal multiplexer improved significantly through seeded TO, exceeding previously reported TO-designed \cite{frellsen2016topology,hiesener2025seeded} and parameter-optimized \cite{uematsu2012design,chack2020broadband,gonzalezandrade2020experimental,guo2017silicon} TE modal multiplexers, and the multiplexer is the largest device optimized with this TO framework to date. The polarization splitter demonstrated, in simulation, a significant improvement in TE\textsubscript{00} transmission, particularly at the long-wavelength edge of the O-band, while slightly improving TM\textsubscript{00} transmission and reducing TM\textsubscript{00} crosstalk, with TE\textsubscript{00} crosstalk remaining roughly the same. The close alignment between simulation and measurement for the fabricated devices indicates strong fabrication robustness for devices designed using this design methodology.

\par

We then applied this validated pipeline to two GF Fotonix\textsuperscript{TM} PDK-provided MMI-based devices, a 1$\times$2 and a 2$\times$2 splitter (Section~\ref{sec_pdk_splitters}); see also \cite{hiesener2026enhancing} for related preliminary results. Seeded TO improved the measured performance of both devices relative to the PDK variant, significantly improving the transmission of the 1$\times$2 splitter and the splitting ratio of the 2$\times$2 splitter, with a mean measured splitting ratio within 0.008 of 0.5 across the measurement band. Simulated over-/under-etch studies performed on each device further demonstrate improved robustness to fabrication variation for the seeded TO variants relative to the PDK devices. 

\par

An optimization paradigm utilizing seeded TO offers many benefits compared to traditional TO design pipelines, including reduced hyperparameter tuning, better-behaved optimization trajectories, and the ability to combine it with another, more well-established design method. We demonstrated the capability of a seeded TO-enhanced design paradigm to design routing, mode-control, and polarization-control devices. The performance of some of the seed devices used (primarily the modal multiplexer) is limited by the accuracy and computational-time cost of the FDTD simulation; without this constraint, the multimode section lengths and widths could be increased, allowing more modes to be excited and improving mode confinement at the interference points. Additional parameters, such as taper location and size, could also be included to excite different modes in each multimode section. Going forward, we are investigating methods to simulate and optimize longer integrated photonic structures that are not feasible to design using conventional FDTD simulations, including cascaded (sectional) simulation, alternative solvers, and GPU-accelerated FDTD.

\par

Because seeded TO is material agnostic, it is also directly applicable to the silicon nitride (Si$_3$N$_4$) waveguide layers available on many foundry platforms, where inverse design has already been demonstrated \cite{song2024ultracompact,slaby2022compact}. The lower index contrast of Si$_3$N$_4$ results in analytically designed devices, such as MMIs, that are longer and have larger features than their silicon counterparts. A Si$_3$N$_4$ seeded TO pipeline would therefore require proportionally larger blurring filters to effectively perturb the seed topology, and would benefit from the same efficient simulation methods being developed for long MMI-based devices. Other investigations include design methods based on evanescent coupling, such as directional couplers \cite{yariv1973coupledmode} and interlayer couplers \cite{li2023highefficiency}. By enabling the enhancement of traditional, well-established devices without requiring a redesign from scratch, seeded TO offers a practical first step toward a TO-enhanced foundry PDK \cite{giewont2019300mm}.

\begin{backmatter}

\bmsection{Funding}
National Science Foundation Center for Electronic-Photonic Integrated Circuits for Aerospace (EPICA) (2052808).

\bmsection{Acknowledgments}
This research was supported in part through research cyberinfrastructure resources and services provided by the Partnership for an Advanced Computing Environment (PACE) at the Georgia Institute of Technology. The authors thank GlobalFoundries for providing silicon fabrication through the Fotonix\textsuperscript{TM} university program. The authors used a generative AI tool (Claude, Anthropic) to assist with editing and proofreading the text of this manuscript. The authors reviewed and edited all AI-assisted content and take full responsibility for the content of this publication.

\bmsection{Disclosures}
The authors declare no conflicts of interest.

\bmsection{Data Availability}
Data underlying the results presented in this paper are not publicly available at this time but may be obtained from the authors upon reasonable request.

\end{backmatter}

\bibliography{refs}

\end{document}